\documentclass[12pt]{article}
\usepackage{times}
\usepackage{geometry}
\usepackage[utf8]{inputenc}
\usepackage{enumitem,amssymb}
\usepackage{ragged2e}
\newlist{thematic}{itemize}{8}
\setlist[thematic]{label=$\square$}
\usepackage{pifont}

\usepackage{graphicx}
\usepackage{amsmath}%
\usepackage{amsfonts}%
\usepackage{amssymb}%
\usepackage{subfig}
\usepackage{wrapfig}
\usepackage{scrextend}
\usepackage{lineno}
\usepackage[sort&compress,numbers]{natbib}
\usepackage{enumitem}
\usepackage{fancyhdr}
\usepackage{xcolor}
\usepackage{ifthen}
\usepackage{ragged2e}
\definecolor{bluecite}{HTML}{0875b7}
\usepackage[unicode=true,
	bookmarksopen={true},
	pdffitwindow=true, 
	colorlinks=true,
	linkcolor=black,
	citecolor=bluecite, 
	urlcolor=bluecite, 
	hyperfootnotes=false, 
	pdfstartview={FitH},
	pdfpagemode= UseNone]{hyperref}
\addtokomafont{labelinglabel}{\sffamily}

\begin{document}

\thispagestyle{empty}

\begin{flushleft}
\huge
Astro2020 Science White Paper \linebreak

\vspace{-0.3cm}

Positron Annihilation in the Galaxy \linebreak
\normalsize

\vspace{-0.15cm}

\noindent \textbf{Thematic Areas:} \hspace*{60pt} $\square$ Planetary Systems \hspace*{10pt} $\square$ Star and Planet Formation \hspace*{20pt}\linebreak
$\square$ Formation and Evolution of Compact Objects \hspace*{31pt} $\boxtimes$ Cosmology and Fundamental Physics \linebreak
  $\boxtimes$ Stars and Stellar Evolution \hspace*{1pt} $\square$ Resolved Stellar Populations and their Environments \hspace*{40pt} \linebreak
  $\square$    Galaxy Evolution   \hspace*{45pt} $\square$             Multi-Messenger Astronomy and Astrophysics \hspace*{65pt} \linebreak

\vspace{-0.15cm}

\textbf{Principal Author:}

Name: Carolyn A. Kierans
 \linebreak						
Institution: NASA/Goddard Space Flight Center
 \linebreak
Email: carolyn.a.kierans@nasa.gov
 \linebreak
Phone: +1-301-286-2597
 \linebreak
 
\vspace{-0.25cm}

\textbf{Co-authors:}
  \linebreak
  John F. Beacom - The Ohio State University \\
  Steve Boggs - University of California, San Diego \\
  Matthew Buckley - Rutgers, The State University of New Jersey \\
  Regina Caputo - NASA/Goddard Space Flight Center \\
  Roland Crocker - Australian National University \\
  Micha\"el De Becker - University of Li\`ege \\
  Roland Diehl - Max Planck Institute for extraterrestrial Physics \\
  Chris L Fryer - Los Alamos National Laboratory \\
  Sean Griffin - NASA/Goddard Space Flight Center/University of Maryland, College Park \\
  Dieter Hartmann - Clemson University  \\
  Elizabeth Hays - NASA/Goddard Space Flight Center \\
  Pierre Jean - Institut de Recherche en Astrophysique et Plan\'etologie \\
  Martin G. H. Krause - University of Hertfordshire \\
  Tim Linden - The Ohio State University \\
  Alexandre Marcowith - Laboratoire Univers et particules de Montpellier \\
  Pierrick Martin - Institut de Recherche en Astrophysique et Plan\'etologie \\
  Alexander Moiseev - NASA/Goddard Space Flight Center \\
  Uwe Oberlack - Johannes Gutenberg University Mainz \\
  Elena Orlando - Stanford University \\
  Fiona Panther - University of New South Wales, Canberra/Australian National University \\
  Nikos Prantzos - Institut d'Astrophysique de Paris \\
  Richard Rothschild - University of California, San Diego \\
  Ivo Seitenzahl - University of New South Wales, Canberra \\
  Chris Shrader - NASA/Goddard Space Flight Center/Catholic University of America \\
  Thomas Siegert - University of California, San Diego \\
  Andy Strong - Max Planck Institute for extraterrestrial Physics \\
  John Tomsick - Space Sciences Laboratory/University of California, Berkeley \\
  W. Thomas Vestrand - Los Alamos National Laboratory \\
  Andreas Zoglauer - Space Sciences Laboratory/University of California, Berkeley \\
\end{flushleft}

\newpage
\pagenumbering{arabic}

\section{Introduction to Galactic Positrons}

The 511~keV $\gamma$-ray line from electron positron annihilation was first detected coming from the Galactic center region in the 1970's with balloon-borne telescopes~\cite{Johnson1972, leventhal1978}. 
After almost 50 years of observations, the origin remains an enigma.
Instruments have measured an extended region with a flux of $\sim$10$^{-3}$~$\gamma$/cm$^2$/s towards the center of the Galaxy~\citep{siegert2016a}, making it the strongest persistent diffuse $\gamma$-ray line. 
The distribution of emission is unlike any other wavelength.
Over the decades, scientists have been aiming to determine the sources of Galactic positrons, but still lack for a clear answer. 
While $\beta^+$-unstable nucleosynthesis products, such as $^{26}$Al and $^{44}$Ti, could possibly explain the disk emission~\citep{prantzos2011}, they fail to account for the excess of positrons in the Galactic Center (GC) region. 
Thus, many Galactic sources have been proposed as the possible birth-site of these positrons: millisecond pulsars \cite{venter2015, Bartels2018_binaries511}, low-mass X-ray binaries \cite{weidenspointner2008,prantzos2006}, neutron star mergers \cite{Fuller2018_511}, stars \cite{Bisnovatyi-Kogan2017_511} and their supernovae (SN, \cite{alexis2014,crocker2017}), pair-plasma jets from Sgr A* that produced the Fermi bubbles \citep{siegert2016a}, and dark matter (DM, \cite{boehm2004b}). 
However, the nature of the source(s) of positrons is still unresolved and highly contested~\cite{Panther18_rev}. 
This remains one of the major puzzles in $\gamma$-ray astrophysics over the last half-century. 

\textbf{The longstanding problem of the unknown source of Galactic positrons could be solved within the next decade.} 
To make progress in the field, the community calls for a wide-field ($>$2~sr), direct imaging, $\gamma$-ray telescope with an all-sky survey mode of operation. 
Specifically, the science requires a telescope with good angular resolution ($\sim$1$^{\circ}$), excellent energy resolution ($\sim$1\% FWHM) within a range of $\sim$200~keV to a few MeV, and at least an order of magnitude improvement in narrow-line sensitivity ($\sim$10$^{-6}$~$\gamma$/cm$^2$/s), with an enhanced sensitivity to low surface brightness emission. 
From an all-sky image with such an instrument, the following measurements can substantially advance our understanding of Galactic positrons:
\begin{itemize}
    \setlength\itemsep{3pt}
    \item map the 511~keV line and positronium continuum over the whole sky with high significance
    \item spatially resolve the annihilation spectra for different regions of the Galaxy
    \item measure the in-flight annihilation spectrum $>$~511~keV to constrain positron injection energy
    \item map the $^{26}$Al 1.8~MeV diffuse emission and the 1.157 MeV line from $^{44}$Ti in SN to understand contributions.
\end{itemize}
These measurements will help constrain positron propagation distances, distinguish between source models, and perhaps determine what proportion of Galactic positrons come from different sources.

The origin of Galactic positrons is not only an interesting puzzle in itself, but determining the true birth sites of the positrons allows the 511~keV emission to be used as a tool to further study the source and environment in question. 
The technology is currently available to make significant progress in our understanding of Galactic positrons, and in turn, increase our understanding of the role of pair plasma in the formation of jets and winds from compact objects~\cite{fender2006, Begelman1984,Beloborodov1999,siegert2016c,Blandford2017}, the escape of cosmic rays in supernova remnants \cite{zirakashvili2011}, and observations can provide tighter constraints for the interaction cross section of the dark matter particle~\cite{sizun2006, beacom2005, beacom2006}.

This paper would like to draw attention to related Astro2020 White Papers: \textit{Catching Element Formation In The Act: The Case for a New MeV Gamma-Ray Mission: Radionuclide Astronomy in the 2020s}~\cite{fryer2020}; \textit{Looking Under a Better Lamppost: MeV-scale Dark Matter Candidates}~\cite{linden2020}; and \textit{Prospects for Galactic Cosmic Rays from Gamma-Ray Observations at MeV and GeV Energies}~\cite{orlando2020}.

\section{Current Observations}

\vspace{0.0in}
\begin{wrapfigure}{R}{3.2in}
\vspace{-0.30in}
\centering
\includegraphics[width=3.2in]{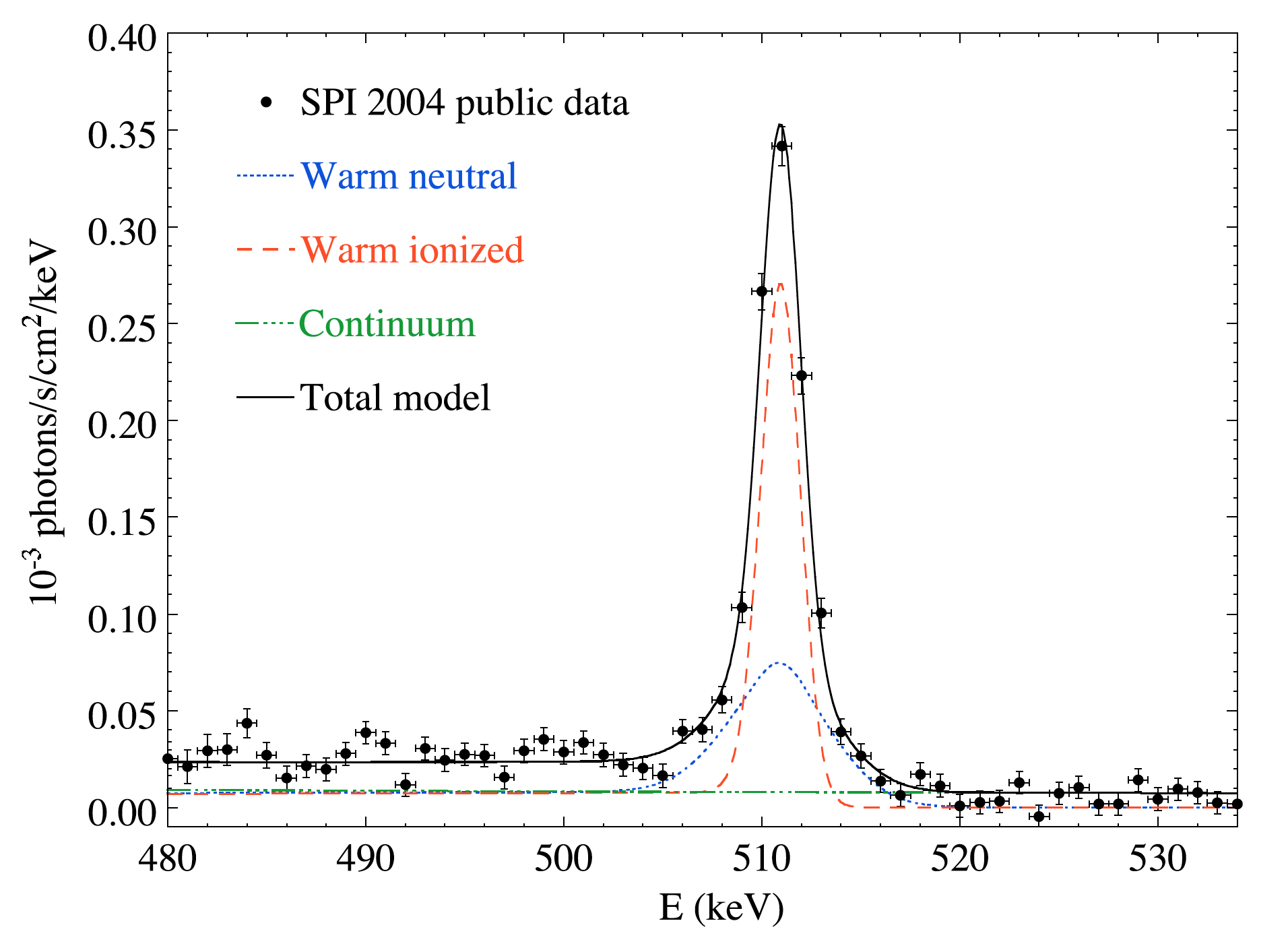}
\vspace{-0.3in}
\caption{\textit{Using 1 year of SPI data, \citet{jean2006} compare the measured 511~keV line profile with derived spectra in different phases of the ISM as calculated in \citet{guessoum2005} to conclude that Galactic positrons predominately annihilate in warm ionized and neutral phases.} \label{fig:jeanspectrum}}
\vspace{-0.0in}
\end{wrapfigure}

The coded-mask spectrometer SPI aboard the \textit{INTEGRAL} satellite~\cite{vedrenne2003} provides the most accurate spectrum of positron annihilation to date. 
In addition to the line at 511~keV, it also precisely measured the three-photon annihilation continuum $<$511~keV from the intermediate bound state of an electron with a positron - positronium~\cite{ore}. 
From measurements of the line-shape as well as the ratio between the 511~keV line and the positronium continuum, it has been determined that $>$95\% of positrons annihilate at low energies ($\sim$eV) compared to their injection ($\gtrsim$MeV, \cite{beacom2005,beacom2006,sizun2006}), and via charge exchange with interstellar gas in warm phases of the ISM~\cite{jean2006}; see Figure~\ref{fig:jeanspectrum}. 
Consequently, positron annihilation can be used as a tool to understand the propagation of low-energy cosmic-rays, as well as the conditions in different phases of the ISM.

\begin{wrapfigure}{L}{3.1in}
\vspace{-0.2in}
\centering
\includegraphics[width=3.1in]{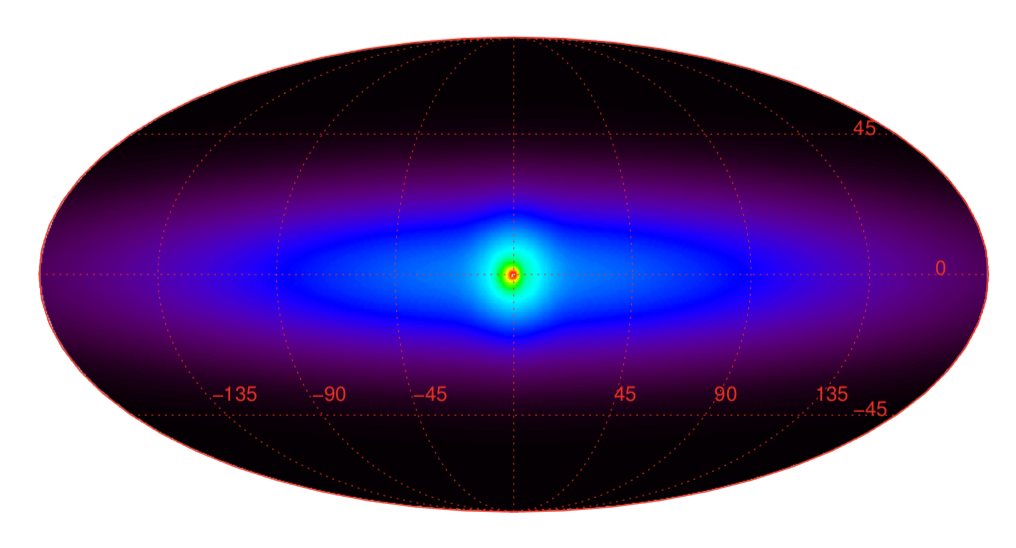}
\vspace{-0.3in}
\caption{\textit{The most recent spatial model derived from INTEGRAL/SPI observations~\cite{siegert2016a}. The central bulge emission is modeled with narrow and broad Gaussian components, with a point source consistent with the GC. The disk emission is fit with a latitude extent of 25$^{\circ}$ FWHM; this is significantly different that the previous thin-disk model by \citet{skinner2014}.}}
\label{fig:spatialmodels}
\vspace{-0.2in}
\end{wrapfigure}

While initial measurements with balloon-borne experiments only detected the bright bulge of the Milky Way in 511~keV emission ~\cite{leventhal1978, albernhe1981}, \textit{INTEGRAL}/SPI was able to measure a contribution from the Galactic disk \cite{skinner2014,siegert2016a}.
As SPI is a coded-mask telescope, the morphology of the diffuse 511~keV emission is derived through a model-fitting approach and remains very uncertain. 
Even after 17 years of data, the bulge shape is not well constrained and the thickness and extent of the low surface brightness disk are not well determined; see Figure~\ref{fig:spatialmodels}.
SPI is not strongly sensitive to structures that are larger than its 16$^{\circ}$ field of view and observations has been strongly concentrated along the Galactic plane; therefore, any high-latitude emission or a possible halo contribution are not easily detectable.

Though SPI has performed spectral characterization of the 511~keV line profile that will be difficult to fundamentally improve upon, the limited ability to measure large scale diffuse emission has left many open questions.
Furthermore, as the total positron annihilation rate depends on the assumed emission morphology, the Galaxy-wide positron annihilation rate is also not well constrained ($2$--$5\times10^{43}$~e$^+$/s).
Therefore, the key to advancing our understanding is a model-independent all-sky image and detailed mapping of characteristic/prominent regions.

\section{Possible Sources of Galactic Positrons}


The $\beta^+$-decay of stellar nucleosynthesis products was first proposed as a primary source of Galactic positrons soon after the initial discovery of the 511~keV line~\cite{clayton1973}, and it still remains the only confirmed contributing source. 
The 1.8~MeV line from the decay of $^{26}$Al, which is produced in large quantities in hydrostatic and explosive nucleosynthesis~\cite{cox1968,woosley1980}, was the second Galactic $\gamma$-ray line to be detected~\cite{Mahoney1984_26Al,pluschke2001}. 
With the measured 1.8~MeV flux, it was found that the $^{26}$Al decay rate corresponds to positron production which is $\sim$10--20\% of the estimated Galaxy-wide rate~\cite{diehl2006}. 
$^{44}$Ti is a second positron-emitting isotope that is released into the ISM by massive stars~\cite{thielemann1996}, but there is more uncertainty in the yields due to unknown SN explosion details~\cite{woosley1995,magkotsios2010}.
The $^{56}$Ni decay chain, which is produced in vast quantities in Type Ia SN, is another possible contributor to the population of Galactic positrons; however, with a limiting half life of 77~days, a major uncertainty is the fraction of positrons which escape the SN explosion region \cite{milne1999}.
$^{26}$Al, $^{44}$Ti, and $^{56}$Ni together can possibly account for the Galaxy-wide rate of positron production; however, the longstanding problem with nucleosynthesis sources is how the observed 511~keV spatial distribution is accounted for considering the dearth of star formation in the bulge.

Microquasars are another proposed source of Galactic positrons that have a long and varied history in the field. 
The ``Great Annihilator'' (1E 1740.7-2942) was observed by SIGMA \cite{sunyaev1991, mirabel1992}; however, the reported claims of 511~keV $\gamma$-rays was later refuted \cite{smith1996}. 
More recently, the detection of a broadened 511~keV line from microquasar V404 Cygni \cite{siegert2016c} has also been controversial \cite{roques2016}. 
In microquasars, positrons can be created through pair-production in the hot inner accretion disk, in the X-ray corona, or at the base of the jet \cite{li1996, paredes2005} and predicted rates of 511~keV emission from the brightest microquasars should be detectable with a next-generation instrument \cite{guessoum2006}.

The spatial distribution of the 511~keV bulge emission is tantalizingly similar to the expected DM distribution in the Milky Way \cite{navarro1997, einasto1965}. 
There are various means of DM positron production \cite{hooper2004, picciotto2005,oaknin2005}, but the most ``natural'' scenarios are direct annihilation of low-mass DM particles ($\lesssim$3 MeV; \cite{beacom2006}) and decay of weakly-interacting massive particles~\cite{pospelov2007}. 
A recent attempt at measuring 511~keV emission from the 40 known Milky Way dwarf galaxies has found a single 3$\sigma$ signal~\cite{siegert2016b}, but a more sensitive instrument is required to constrain the DM contribution of the emission.


A key piece of the puzzle is the unknown distance that positrons propagate between their point of production at MeV energies~\cite{beacom2006,sizun2006} until their annihilation at $\sim$eV~\cite{jean2006}. 
The question of propagation has been the subject of numerous studies with various assumptions and level of computing \cite{jean2009, higdon2009, martin2012, alexis2014, Panther2018_511}, and the details remain elusive.
The most detailed simulations to date find that positrons travel on average $\sim$1~kpc and thus the morphology of the 511~keV emission should be strongly correlated with the source distribution~\cite{alexis2014}. 
As a result, the authors conclude nucleosynthesis alone cannot be the main source of Galactic positrons as the simulated spatial distribution is inconsistent with the SPI images.
The question of the positrons propagation remains highly debated and will benefit from an improved image of the 511~keV emission.

There are also intriguing similarities between the 511~keV emission and the GC excess at GeV $\gamma$-ray energies ($>$10$^6$ keV) as observed by the \textit{Fermi}-Large Area Telescope (LAT)~\cite{Calore2016, Ajello2016}.  
The GeV excess and the bulge component of the 511~keV emission have roughly comparable spatial morphologies;  they are both roughly spherically symmetric, extend about 5--10$^{\circ}$ about the GC, are consistent with a uniform spectrum, and peak strongly toward the GC~\cite{Weidenspointner2006}. 
These similarities suggest a common origin which has been explored by different theoretical models~\cite{gonthier2018,Finkbeiner2007,crocker2017}.


\pagebreak

\section{Potential Results in the Next Decade}

A next-generation, wide-field ($>$2~sr) imaging telescope with an all-sky observing strategy and excellent energy resolution ($\sim$1\%) is needed to progress our understanding of Galactic positrons. 
Though Compton telescopes have an inherent limit on the angular resolution of $\sim$1$^{\circ}$ at 511~keV~\cite{zoglauer2003}, the technology provides a more direct imaging technique and better signal-to-noise than a coded-mask imager. 
Since the launch of \textit{CGRO}/COMPTEL in 1991~\cite{schonfelder1993}, significant progress has been made in Compton telescope technology~\cite{schonfelder2004, boggs2000, kanbach2004}, data post-processing~\cite{zoglauer2006}, and imaging techniques \cite{wilderman1998, zoglauer2011}. 
There are now Compton telescopes in different mission classes and various stages of design and development that can significantly improve upon our current understanding of the MeV sky and positron annihilation, in particular: 
\begin{description}[align=right,labelwidth=3cm, nosep]
    \item [AMEGO]- probe-class concept All-sky Medium Energy Gamma-ray Observatory \\
   \hspace*{2cm} {\footnotesize{\textcolor{bluecite}{\url{https://asd.gsfc.nasa.gov/amego}}}}
    \item [COSI-SMEX]- SMEX-class Compton Spectrometer and Imager \\
    \hspace*{2cm} {\footnotesize{\textcolor{bluecite}{\url{http://cosi.ssl.berkeley.edu}}}}
\end{description}
The AMEGO 
concept is a large scale mission with broad science goals, but nonetheless is well suited for diffuse imaging of $\gamma$-ray lines. 
The success of the COSI balloon payload funded through the APRA program, which is competitive despite its smaller size due to its excellent energy resolution, proves the technology for progress is ready and available~\cite{kierans2016,kieransthesis2018}.

\begin{figure}[t]
    \centering
    \includegraphics[width=0.8\textwidth]{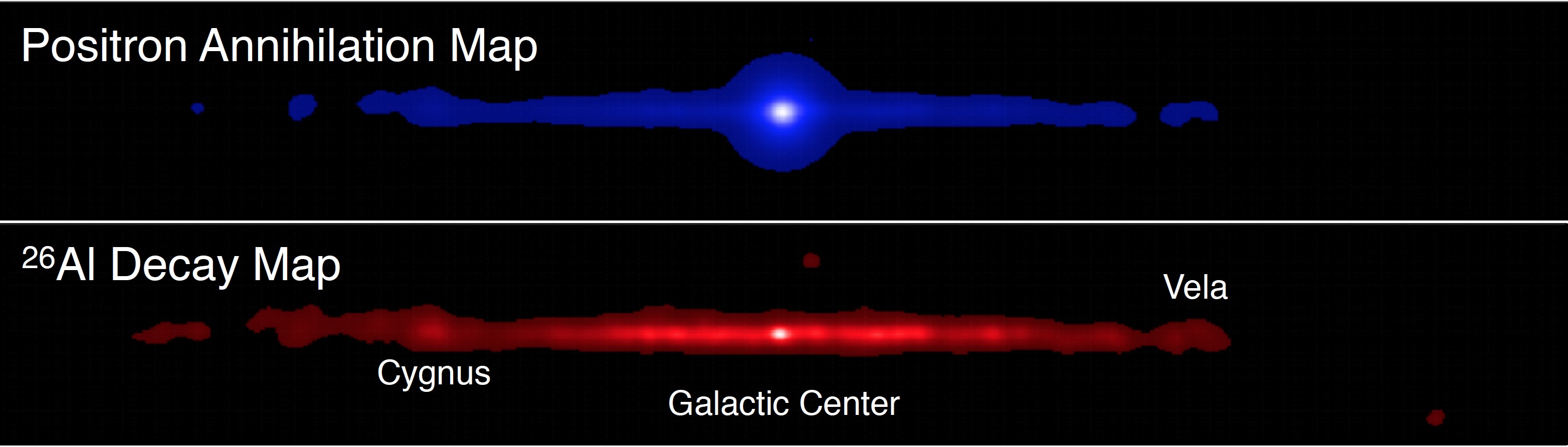}
    \caption{\textit{Simulated imaging results of the entire Galactic plane ($l = \pm180^{\circ}$, $b = \pm 25^{\circ}$) for 2-years of an AMEGO mission. The positron map is a combination of the SPI model from \citet{skinner2014} and the expected contribution from nucleosynthesis based on the COBE DIRBE 240~$\mu$m emission map. The $^{26}$Al decay map also uses 240~$\mu$m as a tracer. 
    The sensitivity of AMEGO was calculated through full Monte-Carlo simulations in MEGAlib~\cite{zoglauer2006} for a NuStar-like orbit.}}
    \label{fig:AMEGOImage}
\end{figure}

Figure~\ref{fig:AMEGOImage} shows the expected spatial distribution of the 511~keV emission and the related $^{26}$Al 1.8~MeV distribution as measured by AMEGO after 2-years of the mission. 
In this image, the 511~keV emission is a combination of the bulge model from \citet{skinner2014} and a disk based on the COBE DIRBE 240~$\mu$m map which is used as a tracer for nucleosynthesis.

\textbf{What are the main sources of Galactic positrons? Is the 511~keV emission truly diffuse? Is there a halo component?} 
As is suggested by Figure~\ref{fig:AMEGOImage}, a lot can be learned from a direct all-sky image of the emission. 
Though improved resolution may help to find point sources and better map the possible fine-structure emission in the Galactic center region, an all-sky map with $\sim$1$^{\circ}$ degree resolution still can address many of the open science questions. 
A Compton imaging telescope could constrain the disk emission height, confirm whether the morphology of the annihilation signal is truly diffuse, and potentially reveal new structures, which then leads to better understanding of possible source contributions. 
There is thought to be a low-surface-brightness halo component, which would change the total positron annihilation rate significantly, and regardless of the angular resolution, with an all-sky observing strategy this contribution could be constrained. 
Additionally, an improved continuum sensitivity $<$511~keV will allow for an image of the positronium spectral component with much higher statistics than the annihilation line.

\textbf{How far do positrons travel before annihilation? How much do nucleosynthesis sources contribute to the population of Galactic positrons?} 
As the decay of $^{26}$Al is known to contribute to the Galactic population of positrons, comparing the observed spatial distribution of the 1.8~MeV $\gamma$-ray line and the mapped 511~keV distribution in specific $^{26}$Al hotspots, such as Cygnus, Scorpius-Centaurus, or Orion-Eridanus, will provide direct observational constraints on the propagation of positrons in the ISM. 
In addition, the evaluation of positron transport properties in isolated regions of the ISM allows for an investigation of low energy particle propagation and ISM dynamics~\cite{padovani2009}. 
For example, the 1.8~MeV line has been detected from the Cygnus complex~\cite{martin2010} and one would expect a corresponding 511~keV flux of $\sim$6$\times 10^{-5}$~$\gamma$/cm$^2$/s in a $\sim$10$^{\circ}$ region if the positrons annihilate quickly; the emission would be more extended if positron propagate. 
In addition, the current maps of $^{26}$Al from COMPTEL~\cite{pluschke2001} and SPI~\cite{bouchet2015} can be improved with any of the above listed telescopes (narrow-line sensitivity $\sim$10$^{-6}$~$\gamma$/cm$^{2}$/s at 1.8~MeV) which will allow for a better direct comparison. 
Likewise, a map of the $^{44}$Ti 1.157~MeV emission along the Galactic plane 
could shed light on the positron contribution from $^{44}$Ti decay without relying on the uncertainties in SNe explosions and yields.

\textbf{How is the positron annihilation environment different throughout the Galaxy?} 
Comparing the spectral line shapes and positronium continuum in different regions of the Galaxy can lead to further understanding of the annihilation conditions, the phase of the surrounding ISM, and could potentially lead to information about different sources of positrons.
An instrument with excellent energy resolution ($<$1\% FWHM) would be needed to measure absolute line shapes, but with a more moderate energy resolution ($\sim$3\%) one can still learn from the relative ones, for example, in the bulge region versus the disk. 
This has been attempted with SPI~\cite{siegert2016a}, but without direct imaging it is difficult to disentangle the emission.


\textbf{What is the positron injection energy? What is the upper limit on the mass of a possible contributing light DM particle?}
A 1-2 order of magnitude improvement in the continuum sensitivity between 511~keV and a few MeV with a next generation instrument will allow one to measure the in-flight annihilation spectrum, which would reveal the birth spectrum of positrons as they are injected into the ISM~\cite{beacom2006}.
The limits on initial energy are currently set by COMPTEL and SPI and have eliminated positron source candidates, e.g. proton-proton interactions in cosmic rays~\cite{porter2008}, and could differentiate between pair-production or nucleosynthesis sources.
This measurement also defines the upper-limit of a possible light DM particle that is considered as a possible source of Galactic positrons~\cite{beacom2005,beacom2006,sizun2006}.

The next decade is the opportune time to have new observations of Galactic positron annihilation. 
The knowledge of the ISM and high energy pair producers like
microquasars will soon be significantly improved thanks to radio (SKA, ngVLA) and X-ray surveys (eROSITA). Gaia has provided a much better grasp on stellar populations, and high-energy sources and compact objects have been scrutinized like never before (\textit{Fermi}, CTA, LIGO/Virgo). 
Solving the positron mystery with better imaging data could add important new constraints that would not only help with our understanding of nucleosynthesis, but also constrain the physics of supernova explosions, the formation of black-hole jets and the properties of dark matter. 

\newpage
\renewcommand{\thepage}{}
\bibliographystyle{yahapj}
\bibliography{references}

\end{document}